# Translation of Quantum Circuits into Quantum Turing Machines for Deutsch and Deutsch-Jozsa Problems

**Giuseppe Corrente**

Università di Torino, Computer Science Department, Via Pessinetto, 12, 10149 Torino,Italy
Corresponding Author: Giuseppe Corrente. Email: giuseppe.corrente@unito.it
Received: XX Month 202X; Accepted: XX Month 202X

**Abstract:** We want in this article to show the usefulness of Quantum Turing Machine (QTM) in a high-level didactic context as well as in theoretical studies. We use QTM to show its equivalence with quantum circuit model for Deutsch and Deutsch-Jozsa algorithms. Further we introduce a strategy of translation from Quantum Circuit to Quantum Turing models by these examples. Moreover we illustrate some features of Quantum Computing such as superposition from a QTM point of view and starting with few simple examples very known in Quantum Circuit form.

## 1 Introduction

One of the first step in the study of Quantum Computing is the Deutsch algorithm for its simplicity, and generally the next step is the study is the Deutsch-Jozsa algorithm for its simplicity and powerful at the same time. Traditionally both, as all most common quantum algorithm, are presented in Quantum Circuit form. This is a starting point also of our article, but we also explore a QTM that solves the Deutsch problem showing so an example of a translation from Quantum Circuit and Quantum Turing Machine models. Finally we do the same for Deutsch-Jozsa problem.

QTM generally is considered an invaluable computational model in the study of the complexity, but not in the study of algorithms. We give an hint in this paper that QTM may be a good model also in this latter area of study, in particular to better understand and illustrate some features of quantum algorithms and some concepts related to the equivalence of different computational models.

## 2 Related works

QTM was introduced for the first time by Deutsch in 1985[1], but only in 1997 QTM and their features are well formalized[2].

Many papers[3][4] and books[5][6] concern about the writing and explanation of note algorithms in Quantum Circuit notation or in some dedicated programming language, but almost no one regard their implementation in a Quantum Turing Machine.

Further [7][8] offer a demonstration of equivalence of QTM and Quantum Circuit families by a general methodology of translation between them, but without showing some translation like that which is the matter of this article.

In [9] a good landscape of different quantum computing models as well as quantum languages is illustrated.

**3 The Deutsch problem**

One of the simplest quantum algorithms is Deutsch's algorithm, his study is a initial obliged step for the Quantum Computing students. This algorithm is about the functions from the set {0, 1} to the set {0, 1}. There are 4 such functions that we can describe as follows:

- x ϵ{0,1}→0
- x ϵ{0,1}→1
- x ϵ{0,1}→ x
- x ϵ{0,1}→ NOT(x)

Note that the third function may would be written also I(x), where I is the identity function.

A function $f : \{0, 1\} \to \{0, 1\}$ is balanced iff $f(0)$ is different from $f(1)$, i.e., it is one to one, and reversible. Differently, a function is constant, that is $f(0) = f(1)$. The four functions above include two balanced and two constant functions. Deutsch's algorithm solves the following problem: Given a function $f : \{0, 1\} \to \{0, 1\}$ as a black box, where one can evaluate an input, but having no knowledge about how the function is defined, he has to determine if the function is constant or not. With a classical computer, one would have to first evaluate $f$ on an input, then evaluate $f$ on the second input and then compare the outputs: with a classical computer, $f$ must be evaluated twice. A quantum computer can be in two states at one time. We shall use this superposition of states to evaluate both inputs at one time. We can show one particular input to feed the black-box-function in a way that obtained output give us the complete answer.

**4 Starting with a quantum gate for a one step evaluation**

Let $f$ the function to evaluate, then the following black-box $U_f$ will be the corresponding quantum gate :

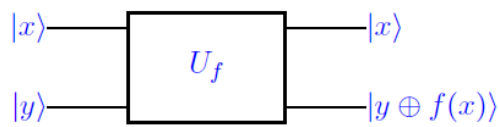

**Figure 1** Quantum gate for f evaluation

The top input, $|x\rangle$, will be the qubit value that we want to evaluate and the bottom input, $|y\rangle$, is a control qubit for the output.

The top output will be the same as the input qubit that is the identity function for that qubit, so to preserve the reversibility, that is a must for quantum gates. The bottom output will be the qubit $|y \oplus f(x)\rangle$ where $\oplus$ is XOR, the exclusive-or operation (binary addition modulo 2). Briefly we can represent the quantum gate for *f* as :

$$U_f : |x, y\rangle \rightarrow |x, y \oplus f(x)\rangle \tag{1}$$

And , obviously, if we apply twice it, we obtain the identity gate, as the reader can check quickly applying $U_f$ twice to initial inputs and taking in mind the associative and idem-potency properties of XOR operator. In the following Figure we can see as $U_f U_f = I$.

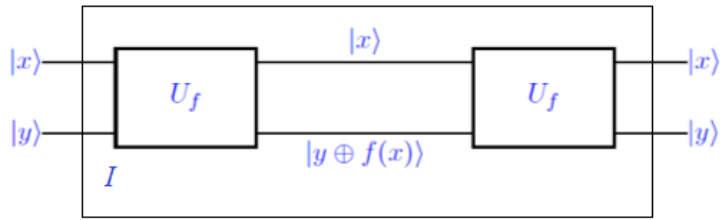

**Figure 2** $U_f$ is unitary

Let us take a first view at a quantum algorithm to solve this problem [10]. Rather than evaluating *f* twice, as in classical solution, we shall try to take an advantage of superposition of states. Instead of having the top input to be either in state $|0\rangle$ or in state $|1\rangle$, we shall put the top input in state $\frac{|0\rangle+|1\rangle}{\sqrt{2}}$ which is the superposition equally balanced as amplitudes of both. The Hadamard gate can place a qubit in such a state, in fact $H|0\rangle = \frac{|0\rangle+|1\rangle}{\sqrt{2}}$. Thus we have the following quantum circuit:

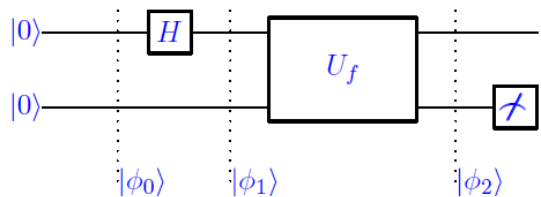

**Figure 3** Evaluating two values in one step

In matrix representation this circuit corresponds to $U_f(H \otimes I)(|0,0\rangle)$. Here H is the Hadamard gate and I is the Identity gate, that in the figure is omitted. We examine the states of the system at every step time. In $|\Phi_1\rangle$ we have the state $\frac{|0,0\rangle+|1,0\rangle}{\sqrt{2}}$. Then, after the application of $U_f$ we have the final state in $|\Phi_1\rangle$ as $\frac{|0,f(0)\rangle+|1,f(1)\rangle}{\sqrt{2}}$. Measuring the bottom qubit we will have with the same probability as result *f(0)* and *f(1)*, loosing so the advantage of quantum superposition.

## 5 Deutsch Algorithm

Now use the previous lesson to actually give Deutsch's algorithm. Deutsch's algorithm works by setting both the top and the bottom qubits into a superposition again using the Hadamard gates. We will also apply an Hadamard gate on the results of the top qubit.

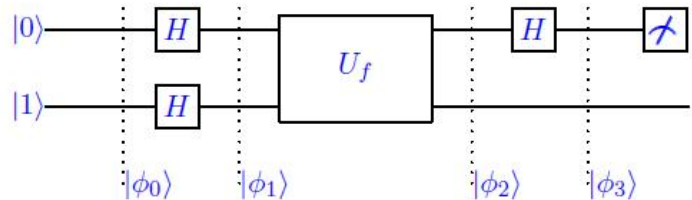

**Figure 4** Quantum Circuit for Deutsch algorithm

In terms of matrices this can be written as $(H \otimes I)U_f(H \otimes H)(|0,1\rangle)$. Here too $H$ are the Hadamard gates and $I$ is the Identity gate, that in the figure is omitted.

Given that $H|0\rangle = \frac{|0\rangle+|1\rangle}{\sqrt{2}}$ and $H|1\rangle = \frac{|0\rangle-|1\rangle}{\sqrt{2}}$, in $|\Phi_1\rangle$ we have the state $\frac{|0,0\rangle-|0,1\rangle+|1,0\rangle-|1,1\rangle}{\sqrt{2}\sqrt{2}}$.

So we have in $|\Phi_2\rangle$, applying the $U_f$ gate, the following state:

$$\frac{((-1)^{f(0)}|0\rangle+(-1)^{f(1)}|1\rangle)(|0\rangle-|1\rangle)}{\sqrt{2}\sqrt{2}} \qquad (2)$$

That the reader can easily verify is equivalent to:

$$\pm \begin{cases} \frac{(|0\rangle+|1\rangle)(|0\rangle-|1\rangle)}{\sqrt{2}\sqrt{2}} & \text{if } f = 0 \text{ or } f = 1 \text{ (costant)} \\ \\ \frac{(|0\rangle-|1\rangle)(|0\rangle-|1\rangle)}{\sqrt{2}\sqrt{2}} & \text{if } f = I \text{ or } f = NOT \text{ (balanced)} \end{cases} \qquad (3)$$

Given that $H\frac{|0\rangle+|1\rangle}{\sqrt{2}} = |0\rangle$ and $H\frac{|0\rangle-|1\rangle}{\sqrt{2}} = |1\rangle$, we have in $|\Phi_3\rangle$, applying the $H$ gate on the top qubit, the following state, omitting the phase :

$$\begin{cases} |0\rangle\frac{(|0\rangle-|1\rangle)}{\sqrt{2}} & \text{if } f = 0 \text{ or } f = 1 \text{ (costant)} \\ \\ |1\rangle\frac{(|0\rangle-|1\rangle)}{\sqrt{2}} & \text{if } f = I \text{ or } f = NOT \text{ (balanced)} \end{cases} \qquad (4)$$

Clearly now a measure on the first qubit solves the problem!

The reader should notice the fact that the output of the top qubit of $U_f$ should not change from being the same as the input. This is not the case because of the inclusion of the Hadamard matrices. This is the essence of the fact that the top and the bottom qubits are both part of the same input. We gain access to the information, but not create it from the null. There are four possible functions, and we know that with a classical computer we needed two bits of information to determine which of the four functions we were given. The core idea of the Deutsch algorithm is to transform data in such a way to answer to the question "Is the function balanced or constant?" and not to the question "What is the value of the function on 0?" or the question "What is the value of the function on 1?" as the direct way does. But, given in this way, the measure done on the top qubit answers to the first of the three above questions, and so it determinates which group among two groups the function belongs.

Let us generalize the Deutsch to the Deutsch-Jozsa algorithm extending the range of functions taken in consideration. Instead of regarding about functions $f : \{0, 1\} \rightarrow \{0, 1\}$, we want manage functions with a larger domain. Consider functions $f : \{0, 1\}^n \rightarrow \{0, 1\}$ (the domain might be thought of as any natural number from $0$ to $2^n - 1$ ). We shall call a function $f : \{0, 1\}^n \rightarrow \{0, 1\}$ balanced if exactly half of the inputs go to 0 (and the other half go to 1). We can call a function constant if all the inputs go to 0 or all the inputs go to 1. The problem solved by Deutsch–Jozsa algorithm is the following: suppose you are given a function from $\{0, 1\}^n$ to $\{0, 1\}$ which you can evaluate without knowing the way it is defined, given before that you are assured that each function considered is either balanced or constant. Our aim is to determine if the function is balanced or constant. Notice that when $n = 1$, this is exactly the problem that the Deutsch algorithm solved. Classically, this algorithm can be solved by evaluating the function on different inputs. The best case scenario is when the first two different inputs have different outputs, which assures us that the function is balanced. In contrast, to be sure that the function is constant, one must evaluate the function on more than half the possible inputs. Different would be the case of a probabilistic algorithm [11]. If the top qubit line in Deutsch-Jozsa algorithm is generalized with to n qubit lines, representing the domain of the new group of functions, and this change is propagated coherently to the whole quantum circuit, we have the Deutsch-Jozsa algorithm starting from the Deutsch algorithm, in quantum circuit notation.

**6 A Quantum Turing Machine For Deutsch algorithm**

We give the following definition of a *Quantum Turing Machine*.

A *Quantum Turing Machine* (QTM for short) M is (Q, Σ , Γ, δ, $q_0$, ⊡ , F) where

- Q is the (finite) set of internal states $\{q_i | i \in N\}$
- Σ is the input alphabet, Γ is the finite set of symbols called tape alphabet (i.e. Γ ∪ ⊡)
- δ : Γ × Q × Γ × Q × {L,N,R} → $C_{[0,1]}$ is the transition function that gives the amplitude of each step, its square represents the probability of that step if a measurement occurs. The range of the transition function are the complex with module equal or less than 1. L,N,R are the moves of the head on the tape admitted, where N indicates that also no head move is admitted
- ⊡ is the blank symbol
- $q_0$ (is a member of Q) is the initial state
- F (is a subset of Q) is the set of final states (one final state is sufficient ).

Follow some other features with the aim to define and explain the meaning of configuration and computation :

- A tape is a pair of strings $w_L$ and $w_R$ such that $w_L \in \square^\infty \Gamma^*$ and $w_R \in \Gamma^* \square^\infty$.
- $h \in \Gamma$ is the head of the tape whenever is the rightmost symbol of $w_L$.
- A configuration is a triple in $Q \times (\square^\infty \Gamma^*) \times (\Gamma^* \square^\infty)$.
- The initial configuration is $<q_0, \square^\infty w, \square^\infty>$ where $w \in \Gamma^*$ is the input.
- An final configuration is $<q_F, \square^\infty w, \square^\infty>$ where $q_F \in F$ and $w \in \Gamma^*$ is the output.
  We assume that if a final configuration is the superposition of more than one then all them are in a final state.

- A TM-computation is a (finite) sequence of configurations $c_0, \ldots, c_n$ such that, for all $i \in [0, n-1]$, if $c_i = <q_i, w_L^i:h, w_R^i>$ then $c_{i+1}$ is obtained by applying the transition rule in the straightforward way, i.e. by respecting the information of $\delta(q_i, h) = [q_{i+1}, h_{i+1}, m]$, so if this rule is respected we can affirm that the computation is deterministic, and the range of transition function becomes $\{0,1\}$ from $C_{[0,1]}$. Note that in this case the transition function can be properly represented as $\delta : \Gamma \times Q \times \Gamma \times Q \times \{L,N,R\} \to \{0,1\}$, or as $\delta : \Gamma \times Q \to \Gamma \times Q \times \{L,N,R\}$ equivalently cleaning totally all tuples going to 0 in the first representation.
- A QTM-computation[12] is a (finite) set of configurations $C_M$ above which $\delta$ determines a mapping $a: C_M \times C_M \to C_{[0,1]}$ such that for each $c_1, c_2 \in C_M \times C_M$, $a(c_1, c_2) \in C_{[0,1]}$ represents the amplitude of the transition of M from $c_1$ to $c_2$. This matrix has to be unitary.

So, for each configuration $c_0$ and all its successor configurations $c_1, \ldots, c_k$ the following sentence must be true: if $\alpha_i$ is the amplitude assigned to the transition from $c_0$ to the configuration $c_i$, then

$$\sum_{i=1}^{k} |\alpha_i|^2 = 1 \qquad (5)$$

where $|\alpha_i|^2$ represents the probability of transition from $c_0$ to $c_i$, but $c_1, \ldots, c_k$ occur in parallelism and so step by step until a measurement is not effectuated.

Note that after the first step the starting configuration $c_0$ can be also a superposition of configurations, in this case the next configuration is always determined by the transition function, but weighting each component of $c_0$ with the relative amplitude.

We will represent the transition function with a matrix adding some conditions, with the aim to represent the Deutsch algorithm with a QTM . Referring to Deutsch algorithm we assume having a single tape with two symbols representing initial top and bottom qubits.

Let

- $Q = \{\Phi_0, \Phi_1, \Phi_2, \Phi_3\}$
- $\Sigma = \{00, 01, 10, 11, \square\}$
- $q_0 = \Phi_0$
- $F = \{\Phi_3\}$

then if we define δ with the following rules, we have a QTM that performs the Deutsch algorithm. Note that these are written taking in mind the relative circuit design and the operating principles of Hadamard (H) gate.

$$\delta\,(\square, \Phi_0, 01, \Phi_0, N) = 1 \tag{6}$$

$$\delta\,(01, \Phi_0, 00, \Phi_1, N) = 1/2 \tag{7}$$
$$\delta\,(01, \Phi_0, 01, \Phi_1, N) = -1/2 \tag{8}$$
$$\delta\,(01, \Phi_0, 10, \Phi_1, N) = 1/2 \tag{9}$$
$$\delta\,(01, \Phi_0, 11, \Phi_1, N) = -1/2 \tag{10}$$

$$\delta\,(00, \Phi_1, 0f(0), \Phi_2, N) = 1 \tag{11}$$
$$\delta\,(01, \Phi_1, 0\ \mathrm{NOT}(f(0)), \Phi_2, N) = 1 \tag{12}$$
$$\delta\,(10, \Phi_1, 1f(1), \Phi_2, N) = 1 \tag{13}$$
$$\delta\,(11, \Phi_1, 1\ \mathrm{NOT}(f(1)), \Phi_2, N) = 1 \tag{14}$$

$$\delta\,(0x, \Phi_2, 0x, \Phi_3, N) = 1/\sqrt{2} \tag{15}$$
$$\delta\,(0x, \Phi_2, 1x, \Phi_3, N) = 1/\sqrt{2} \tag{16}$$
$$\delta\,(1x, \Phi_2, 0x, \Phi_3, N) = 1/\sqrt{2} \tag{17}$$
$$\delta\,(1x, \Phi_2, 1x, \Phi_3, N) = -1/\sqrt{2} \tag{18}$$

The reader should observe that no head move is present in the rules and the initial qubits are generated from blank symbol in the initial rule. The real input is not given on the tape but by the function *f* embedded in the rules 11,12,13 and 14.

Note that we can substitute superposition among rules, represented apparently by indeterminism, by superposition on the tape. For example rules from 7 to 10 may be substituted with the following couple of rules:

$$\delta\,(01, \Phi_0, \tfrac{1}{\sqrt{2}}(|0\rangle + |1\rangle)0, \Phi_{1,}, N) = 1/\sqrt{2} \tag{7a,9a}$$
$$\delta\,(01, \Phi_0, \tfrac{1}{\sqrt{2}}(|0\rangle + |1\rangle)1, \Phi_{1,}, N) = -1/\sqrt{2} \tag{8a,10a}$$

or equivalently with the following couple of rule

$$\delta\,(01, \Phi_0, 0\,\tfrac{1}{\sqrt{2}}(|0\rangle - |1\rangle), \Phi_{1,}, N) = 1/\sqrt{2} \tag{7b,8b}$$
$$\delta\,(01, \Phi_0, 1\,\tfrac{1}{\sqrt{2}}(|0\rangle - |1\rangle), \Phi_{1,}, N) = 1/\sqrt{2} \tag{9b,10b}$$

Further, always taking in mind the same principle, they may be substituted by a single equation, expressing all the superposition on the tape instead that by the multiplicity of rules :

$$\delta\,(01, \Phi_0, \tfrac{1}{2}(|0\rangle + |1\rangle)(|0\rangle - |1\rangle), \Phi_{1,}, N) = 1 \tag{7c,8c,9c,10c}$$

Similarly, rules from 15 to 18 may be rewritten in the following form:

$$\delta\ (0x, \Phi_2, \tfrac{1}{\sqrt{2}}(|0\rangle + |1\rangle)x, \Phi_3, N) = 1 \qquad (15d,16d)$$

$$\delta\ (1x, \Phi_2, \tfrac{1}{\sqrt{2}}(|0\rangle - |1\rangle)x, \Phi_3, N) = 1 \qquad (17d,18d)$$

Anyway, independently of the form chosen for the rules, this QTM perfectly simulates the Deutsch algorithm. It starts with a blank tape and in the initial state, writing the fixed symbol 01 on it representing the top and bottom qubit of the corresponding quantum circuit.

The group of rules from 7 to 10 represent applying Hadamard gates to both qubits, and so we advance one step in quantum algorithm state. The rues from 11 to 14, applied to superposition of states obtained by previous step, result in simultaneous calculus of *f(0)* and *f(1)* performed by $U_f$ gate.

Finally the rules from 15 to 18 correspond to apply the Hadamard gate to the top qubit and Identity gate to the bottom qubit in the last step. Now we have reached the final state, and reading on the tape the first symbol we obtain the answer of the Deutsch problem.

## 7 Generalizing the Quantum Turing Machine From Deutsch to Deutsch-Jozsa algorithm

Let

- $Q=\{\Phi_0, \Phi_1, \Phi_2, \Phi_3\}$
- $\Sigma=\{x_1 x_2 \ldots x_n | x_i \in \{0,1\}\} \cup \{\square\}$
- $q_0 = \Phi_0$
- $F = \{\Phi_3\}$

then if we define δ with the following rules, we have a QTM that performs the Deutsch-Jozsa algorithm.

$$\delta\ (\square, \Phi_0, 0^{n-1}1, \Phi_0, N) = 1 \qquad (19)$$

$$\delta\ (0^{n-1}1, \Phi_0, (|0\rangle + |1\rangle)^{n-1}0, \Phi_1, N) = 1/\sqrt{2^n} \qquad (20)$$
$$\delta\ (0^{n-1}1, \Phi_0, (|0\rangle + |1\rangle)^{n-1}1, \Phi_1, N) = -1/\sqrt{2^n} \qquad (21)$$

$$\delta\ (x_1 x_2 \ldots x_{n-1} 0, \Phi_1, x_1 x_2 \ldots x_{n-1} f(x_1 x_2 \ldots x_{n-1}), \Phi_2, N) = 1 \qquad (22)$$
$$\delta\ (x_1 x_2 \ldots x_{n-1} 1, \Phi_1, x_1 x_2 \ldots x_{n-1}\ \text{NOT}(f(x_1 x_2 \ldots x_{n-1})), \Phi_2, N) = 1 \qquad (23)$$

$$\delta\ (x_1 x_2 \ldots x_{n-1} 0, \Phi_2, H(x_1 x_2 \ldots x_{n-1})0, \Phi_3, N) = 1 \qquad (24)$$
$$\delta\ (x_1 x_2 \ldots x_{n-1} 1, \Phi_2, H(x_1 x_2 \ldots x_{n-1})1, \Phi_3, N) = 1 \qquad (25)$$

This QTM perfectly simulates the Deutsch-Jozsa algorithm, the dimostration for induction on n follows.

For n=2 it is equivalent to Deutsch QTM , in fact:

Rule 19 → Rule 6

Rules 20,21 → $\delta (01, \Phi_0, (|0\rangle + |1\rangle)0, \Phi_1, N) = 1/\sqrt{4}$, $\delta (01, \Phi_0, (|0\rangle + |1\rangle)1, \Phi_1, N) = -1/\sqrt{4}$
→ rules 7,8,9,10

Rule 22 → rules 11,13
Rule 23 → rules 12,14

Rule 24,25 → rules 15,16,17,18

Now we write the rules for a QTM for Deutsch algorithm with exactly the same formalism of that for Deutsc-Jozsa putting n=2

$\delta (\square, \Phi_0, 01, \Phi_0, N)=1$ (19')

$\delta (01, \Phi_0, (|0\rangle + |1\rangle)0, \Phi_1, N) = \frac{1}{2}$ (20')
$\delta (01, \Phi_0, (|0\rangle + |1\rangle)1, \Phi_1, N) = -\frac{1}{2}$ (21')

$\delta (x0, \Phi_1, xf(x), \Phi_2, N) = 1$ (22')
$\delta (x1, \Phi_1, xNOT(f(x)), \Phi_2, N) = 1$ (23')

$\delta (x0, \Phi_2, H(x)0, \Phi_3, N) = 1$ (24')
$\delta (x1, \Phi_2, H(x)1, \Phi_3, N) = 1$ (25')

If it solves Deutsch-Jozsa algorithm for $U_f$ with f on *n-1* binary input variables, then its extension by 1 performs the Deutsch-Jozsa algorithm for $U_f$ with f on *n* binary input variables.

So we have:

- $Q=\{\Phi_0, \Phi_1, \Phi_2, \Phi_3\}$
- $\Sigma=\{x_1x_2...x_{n+1}| x_i \in \{0,1\} \} \cup \{\square\}$
- $q_0=\Phi_0$
- $F=\{\Phi_3\}$

And the rules:

$\delta (\square, \Phi_0, 0^n1, \Phi_0, N)=1$ (26)

$\delta (0^n1, \Phi_0, (|0\rangle + |1\rangle)^n 0, \Phi_1, N) = 1/\sqrt{2^{n+1}}$ (27)
$\delta (0^n1, \Phi_0, (|0\rangle + |1\rangle)^n 1, \Phi_1, N) = -1/\sqrt{2^{n+1}}$ (28)

$\delta (x_1x_2...x_n0, \Phi_1, x_1x_2...x_n f(x_1x_2...x_n), \Phi_2, N) = 1$ (29)
$\delta (x_1x_2...x_n1, \Phi_1, x_1x_2...x_n NOT(f(x_1x_2...x_n)), \Phi_2, N) = 1$ (30)

$\delta (x_1x_2...x_n0, \Phi_2, H(x_1x_2...x_n)0, \Phi_3, N) = 1$ (31)
$\delta (x_1x_2...x_n1, \Phi_2, H(x_1x_2...x_n)1, \Phi_3, N) = 1$ (32)

In particular note that 27 and 28 are implied by 20 and 21 rules in consequence of the facts

- $H(y_1 y_2 \ldots y_m) = H(y_1 y_2 \ldots y_{m-1})H(y_m)$ (33)
- $H(0) = (|0\rangle + |1\rangle)/\sqrt{2}$ (34)
- $H(1) = (|0\rangle - |1\rangle)/\sqrt{2}$ (35)

Note also that rule 24 implies 31 and that 25 implies 32 in consequence that :

- $H(y_1 y_2 \ldots y_m) = H(y_1 y_2 \ldots y_{m-1})H(y_m)$ (36)

These deductions together with the induction methodology or the correspondence step by step of QTM Deutsch-Jozsa rules with the relative Quantum Circuit design of this algorithm bring to the proof. So our QTM represents a class of Quantum Turing Machines that solves Deutsch and Deutsch-Jozsa problems for all input of size *n*.

Now we want underline the role of quantum oracle for f in Deutsch and Deutsch-Jozsa algorithm. Both for Deutsch and the Deutsch-Jozsa algorithm we use an $U_f$ gate with x,y input where x=$x_1 x_2 \ldots x_{n-1}$ (n=2 for Deutsch algorithm) that is an n-1 input qubit and y a control qubit for $U_f$. When $x_i$ and y assume only classical values, 0 and 1, $U_f$ works as a standard classical gate. When $x_1..x_{n-1}$,y are considered qubits they may have superposition of values and this permits $U_f$ to calculate more than a value for *f* in one step. So in this case it represents a quantum oracle.

$U_f$ is translated in the QTM for Deutsch-Jozsa in the rules 22 and 23, note that when n=2 these value for Deutsch algorithm. For the QTM definition, in particular the property regarding the transition function when it starts from a superposition of configurations, running these rules on inputs $xy = 1/\sqrt{2^n}(|0\rangle + |1\rangle)^{n-1}(|0\rangle - |1\rangle)$ we have again a quantum oracle, that is they are equivalent to $U_f$. Obviously would be the same using for Deutsch algorithm the four rules in the form 11,12,13,14 instead of the two in the form 22' and 23'.

**Conclusions**

In this article we show that QTM can in some cases be very useful to deepen, understand and illustrate one or some Quantum algorithm starting from circuit representation, converting it and studying in the new representation their relevant features.

We also sketch a proof by induction to further validate Deutsch-Jozsa quantum algorithm by using QTM, and illustrate the features of superposition and quantum oracle, by mean of an example based on the Deutsch and Deutsch-Jozsa algorithm expressed in the QTM model.

**Acknowledgments**

We are grateful to Luca Paolini of Università di Torino for his support and encouragement.

**Funding Statement**

The author(s) received no specific funding for this study.

**Conflicts of Interest**

The authors declare that they have no conflicts of interest to report regarding the present study.